\documentclass[aps,superscriptaddress,showpacs,floatfix]{revtex4}
\usepackage{graphicx}
\usepackage{dcolumn}
\usepackage{bm}
\usepackage{color,soul}
\usepackage{CJK}

\begin{document}

\title{Proton spectral functions in finite nuclei based on the extended Brueckner-Hartree-Fock
approach}

\author{Pei Wang}
\affiliation{National Astronomical Observatories, Chinese Academy of Sciences, Beijing 100012, China}

\author{Peng Yin \footnote{Corresponding author: yinpeng@impcas.ac.cn}}
\affiliation{Institute of Modern Physics, Chinese Academy of
Sciences, Lanzhou 730000, China}\affiliation{Department of Physics and Astronomy, Iowa State University, Ames, IA 50011, USA}

\author{Xinle Shang \footnote{Corresponding author: shangxinle@impcas.ac.cn}}
\affiliation{Institute of Modern Physics, Chinese Academy of
Sciences, Lanzhou 730000, China}

\author{Wei Zuo \footnote{Corresponding author: zuowei@impcas.ac.cn}}
\affiliation{Institute of Modern Physics, Chinese Academy of
Sciences, Lanzhou 730000, China}\affiliation{University
of Chinese Academy of Sciences, Beijing, 100049, China}

\begin{abstract}
We have calculated the proton spectral functions in finite nuclei based
on the local density approximation where the properties of finite nuclei and
nuclear matter are calculated by the Skyrme-Hartree-Fock method
and the extended Brueckner-Hartree-Fock approach, respectively.
The scaled spectral function from our calculation is in good
agreement with experimental results at small momenta while the
difference between them becomes apparent at high momenta.
Besides, a target dependence of the scaled proton spectral
function is also obtained in our calculation as was observed in
experiment. A further investigation indicates that the proportion of the high density region of the proton has a significant contribution to this target-dependent behavior since the spectral function in asymmetric nuclear matter increases significantly as a function of density.
\end{abstract}

\pacs{21.30.Fe, 21.65.+f, 24.10.Cn} \maketitle

\section{Introduction}
Nucleon-nucleon (NN) short-range correlations which are induced by the hard core in the bare NN potential are of great interest since they are closely related to the properties of neutron-rich nuclei, particle production in heavy-ion collisions, as well as neutron star physics~\cite{Schiavilla:2007,Buss:2012,Frankfurt:2008}. It leads to a new challenge to test
 the validity of the physical picture of independent particle motion in the mean field theory
 or the standard shell model~\cite{cavedon:1982,pandharipande:1997}.

 In experiment, the effects of NN correlations can be investigated by
the ($e,e'p$), ($e,e'NN$) and proton induced knock-out reactions~\cite{ramos:1989,dickhoff:1992,dickhoff:2004}.
The related measurements have been reported
continually~\cite{mitt:1990,lapikas:1999,starink:2000,batenburg:2001,rohe:2004,niyazov:2004,
benmokhtar:2005,aclander:1999,onderwater:1998,reley:2008,subedi:2008} and definite evidence of
short-range NN correlations has been observed in these experiments. One important measurement of such medium modifications is the spectral function which can be observed in electron scattering experiments~\cite{Dickhoff:1994}.

Nucleon spectral function in nuclear matter has been calculated by
adopting various many-body methods, such as the relativistic
Dirac-Brueckner-Hartree-Fock  theory~\cite{Dalen:2010}, the
transport model~\cite{Konrad:2005}, the in-medium T-matrix
approach~\cite{Bozek:2004}, the self-consistent Green's function
method~\cite{Frick:2003,Rios:2009,Frick:2005}, and the
Brueckner-Hartree-Fock (BHF)approach~\cite{ Hassaneen:2004}.
Within different approaches, the main features of the spectral
functions turn out to have similar behavior in the region relevant to the
short-range correlations. However, discrepancies of the predicted
spectral functions based on different methods are still present
and controversial interpretations of the experimental data
exist~\cite{ Baldo:2002}. Besides, the information about the
nuclear spectral function and the effects of NN correlations in
finite nuclear systems have also been explored in theory
~\cite{Frick:2004,Muther:1994,Muther:1995a,Benhar:1994,Muther:1995b}.
In Ref.~\cite{Frick:2004}, the authors pointed out that the effects
of short-range correlations are insensitive to the bulk structure
of the nuclear system by a comparison of the spectral function
derived from experimental data with the one obtained from the
Green's function method in nuclear matter, while only the spectral
function of $^{12}$C was reported without a systematic
investigation for different nuclei. Within the framework of
Correlated Basis Function approach, the scaled proton spectral
function, i.e., the proton spectral function scaled by the number
of protons, was compared for different nuclei and a significant
increase of the scaled spectral function with the mass number was
observed. In Ref.~\cite{Rohe:2003}, experimental results of electron scattering on nuclei are available for several nuclei ($^{12}$C, $^{27}$Al, $^{56}$Fe and $^{197}$Au) and  the spectral function was explained as an indication of short-range correlations. In Ref.~\cite{Bozek:2004}, the author calculated the spectral functions of the four nuclei ($^{12}$C, $^{27}$Al, $^{56}$Fe and $^{197}$Au) in the self-consistent T-matrix approach and the target
dependence of the scaled proton spectral function is expected to
stem partially from the asymmetry of target and the density
dependence of the spectral function by using a local density approximation
(LDA). However, the effect of the
density dependence of the spectral function on the target
dependence of the scaled spectral function was not illustrated
explicitly and the reason for the significant difference of
the spectral functions going from $^{12}$C to $^{197}$Au was not
explained. In our previous research, the extended BHF approach has
been adopted to calculate the spectral functions in symmetric as
well as asymmetric nuclear matter, and the three-body force effect
on the spectral function in nuclear matter has been
investigated~\cite{Wang:2013,Wang:2014}. In the present paper, the
extended BHF approach supplemented by the LDA will be firstly
applied to calculate the spectral functions in finite nuclei. One
of our purposes in the present paper is to investigate the strong
target dependence of the scaled proton spectral function within
the framework of the extended BHF approach.

 In the present paper, we shall calculate the scaled spectral function from the LDA~\cite{ Neck:1995}. The calculation for finite nuclei and nuclear matter shall be performed within the framework of the Skyrme-Hartree-fock (SHF) method and  the extended BHF approach, respectively. In this paper, only the spectral functions of the four nuclei, i.e., $^{12}$C, $^{27}$Al, $^{56}$Fe and $^{197}$Au are calculated while similar calculations can be naturally applied and extended to other nuclei.

The present paper is organized as follows. In the next section, we
give a brief review of the adopted theoretical approaches including the extended BHF theory and a microscopic three-body force (TBF) model. We also give definitions and the corresponding physical interpretations of the mass operator and spectral function.
In Section III, the calculated results will be reported and discussed. Finally, a summary is given in Section IV.

\section{Theoretical Approaches}
The present calculations are based on the extended BHF approach
for asymmetric nuclear matter~\cite{zuo:1999}. Here for
completeness, we simply give a brief review about this theory. The
starting point of the BHF approach is to obtain the reaction
$G$-matrix by solving the following isospin dependent
Bethe-Goldstone (BG) equation~\cite{day},
\begin{eqnarray}\label{eq:BG}
G(\rho,\beta;\omega)= V_{NN}+
%\nonumber\\
 V_{NN}\sum\limits_{k_{1}k_{2}}\frac{|k_{1}k_{2}\rangle
Q(k_{1},k_{2})\langle
k_{1}k_{2}|}{\omega-\epsilon(k_{1})-\epsilon(k_{2})}G(\rho,\beta;\omega)
%\nonumber\\
\end{eqnarray}
where $k_i\equiv(\vec k_i,\sigma_i,\tau_i)$ denotes the momentum,
the $z$-component of spin and isospin of a nucleon, respectively.
$Q(k_{1},k_{2})=[1-n_0(k_{1})][1-n_0(k_{2})]$ is the Pauli
operator which prevents two nucleons in intermediate states from being scattered into their
respective Fermi seas (Pauli blocking effect). Here $n_0(k)$
denotes the Fermi distribution function and it is given by a step function at zero temperature, i.e.,
$n_0(k)=\theta(k_F-k)$. The asymmetry parameter $\beta$ is defined as
$\beta=(\rho_{n}-\rho_{p})/\rho$, where $\rho$, $\rho_n$ and $\rho_p$ denote
the total nucleon, neutron and proton
number densities, respectively. $V_{NN}$ is the bare NN interaction and $\omega$ is the starting energy.
$\epsilon(k)$ is the single-particle (s.p.) energy which is given by:
$ \epsilon(k)=\hbar^{2}k^{2}/(2m)+U(k)$.
Here the auxiliary s.p. potential $U(k)$ controls the convergent rate of the
hole-line expansion~\cite{day} and
the continuous choice for
the auxiliary potential is adopted in the present calculation since it provides a much faster convergence of
the hole-line expansion up to high densities than the gap choice~\cite{song:1998}.
 Under the continuous choice, the s.p. potential describes physically at the lowest BHF level
the nuclear mean field felt by a nucleon in nuclear medium~\cite{lejeune:1978} and is
calculated as follows:
\begin{eqnarray}\label{eq:UBHF}
U(k)=Re\sum\limits_{k'\leq k_{F}}\langle
kk'|G[\rho,\epsilon(k)+\epsilon(k')]|kk'\rangle_{A} \ ,
\end{eqnarray}
where the subscript $A$ denotes anti-symmetrization of the matrix
elements.

In the present calculation, we adopt the
Argonne $V_{18}$ (AV18) two-body interaction~\cite{wiringa:1995} plus a
microscopic TBF~\cite{zuo:2002a} constructed by using the meson-exchange current
approach~\cite{grange:1989} for the realistic NN interaction $V_{NN}$.
In the TBF model adopted here, the most important mesons, i.e., $\pi$, $\rho$, $ \sigma $ and $\omega$ have been considered.
The parameters of the TBF model, i.e., the coupling constants
and the form factors,
have been self-consistently determined to reproduce the AV18
two-body force using the one-boson-exchange potential
model and their values can be found in Ref.~\cite{zuo:2002a}.
 In our calculation, the TBF
contribution has been included by reducing the TBF to an
equivalently effective two-body interaction according to the
standard scheme as described in Ref.~\cite{grange:1989}. The extension of the BHF scheme to
include microscopic three-body forces can be found in Ref.~\cite{grange:1989,zuo:2002a,zuo:2002b}. In
$r$-space, the equivalent two-body force $V_3^{\rm eff}$ reads:
\begin{eqnarray}\label{eq:tbf}
 \langle \vec r_1^{\ \prime} \vec r_2^{\ \prime}| V_3^{\rm eff} |
\vec r_1 \vec r_2 \rangle = \displaystyle
 \frac{1}{4} {\rm Tr} \sum_{n} \int {\rm d}
{\vec r_3} {\rm d} {\vec r_3^{\ \prime}}\phi^*_n(\vec r_3^{\
\prime}) (1-\eta(r_{13}')) (1-\eta(r_{23}')) \nonumber \\
\times W_3(\vec r_1^{\ \prime}\vec r_2^{\ \prime} \vec r_3^{\
\prime}|\vec r_1 \vec r_2 \vec r_3)
 \phi_n(\vec r_3)
(1-\eta(r_{13}))
 (1-\eta(r_{23})).
\end{eqnarray}

 Within the framework of the Brueckner-Bethe-Goldstone theory, the mass operator can be expanded in a perturbation series
according to the number of hole lines, i.e.,
\begin{equation}\label{eq:mass}
 M^{\tau}(k,\omega) = M^{\tau}_1(k,\omega)+M^{\tau}_2(k,\omega)+M^{\tau}_3(k,\omega)+\cdots
 \end{equation}
where $\tau$ denotes neutron or proton (hereafter we will write out explicitly the isospin index $\tau$). The mass operator is complex quantity, i.e., $M^\tau(k,\omega)=V^\tau(k,\omega)+iW^\tau(k,\omega)$ and the real part of its on-shell value
can be identified with the potential energy felt by a neutron or a proton
in asymmetric nuclear matter. In the present calculation, we consider the first two terms
$M_1^\tau(k,\omega)$ and $M_2^\tau(k,\omega)$. $M_1^\tau(k,\omega)$ corresponds to the standard BHF s.p. potential. $M_2^\tau(k,\omega)$ is called the Pauli rearrangement term which describes the effects of the ground-state two-hole correlations on the s.p. potential~\cite{baldo:1988,baldo:1990}. The detailed expressions for $M_1^\tau(k,\omega)$ and $M_2^\tau(k,\omega)$ can be found in Refs.~\cite{zuo:1999}.
According to the Lehmann representation for the Green function $G^\tau(k,\omega)=[\omega-k^2/2m-M^\tau(k,\omega)]^{-1}$, the nucleon spectral functions
in nuclear matter can be expressed as follows, i.e.,
\begin{equation}\label{eq:specmatter}
A^\tau(k,\omega)=-\frac{1}{\pi}\frac{W^\tau(k,\omega)}{[\omega-k^2/2m-V^\tau(k,\omega)]^2
+[W^\tau(k,\omega)]^2},
 \end{equation}
 and it fulfils the sum rule of $\int^{\infty}_{-\infty}A^\tau(k,\omega)d\omega=1$.

One of the main purposes of the present work is to investigate the proton spectral functions of the finite nuclei system. To simplify the calculation, we adopt the spherical assumption for
nuclei and the LDA has been applied. The proton spectral function for a finite nucleus is
calculated by the radial integral of the proton spectral function in nuclear matter, and the
expression reads
 \begin{equation}\label{eq:specnuclei}
S^p(k,E)=\frac{2}{(2\pi)^3}\int A^p_h[\rho(r),\beta(r);k,E]d^3\vec r,
 \end{equation}
where $\rho(r)$ and $\beta(r)$ are the local density and isospin asymmetry
at radius $r$, respectively. $A^p_h[\rho(r),\beta(r);k,E]$ is the
proton hole spectral function in nuclear matter. The proton
exciting energy is obtained by $E=\omega_F[\rho(r),\beta(r)]-
 \omega[\rho(r),\beta(r)]|_{\omega<\omega_F}$, where
 $\omega_F$ is the Fermi energy of nuclear matter.

\section{Results and Discussions}

\begin{figure}[tbh]
\begin{center}
\includegraphics[width=0.6\textwidth]{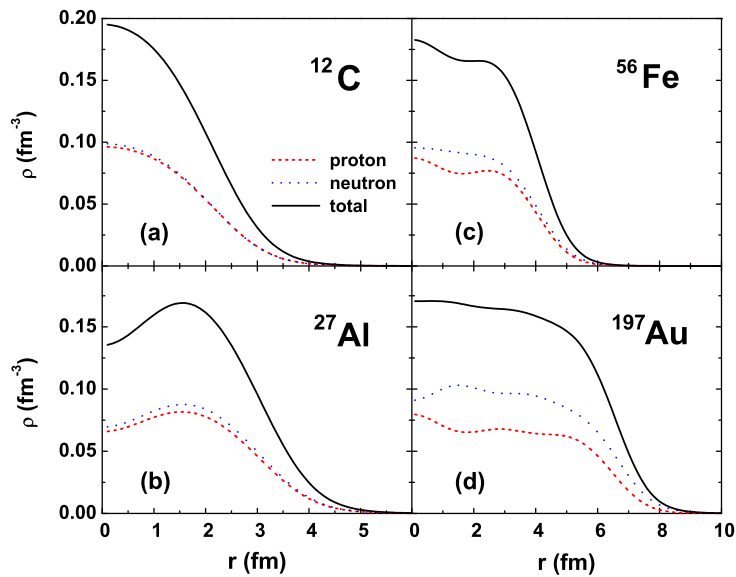}
\end{center}
\caption{Radial density distribution of the four nuclei $^{12}$C (a), $^{27}$Al (b), $^{56}$Fe (c) and
$^{197}$Au (d) calculated by the SHF method with the LNS1 parameter set. The solid lines denote the total
densities while the dashed and dotted lines represent the proton and neutron densities
respectively.}\label{fig1}
\end{figure}

\begin{figure}[tbh]
\begin{center}
\includegraphics[width=0.5\textwidth]{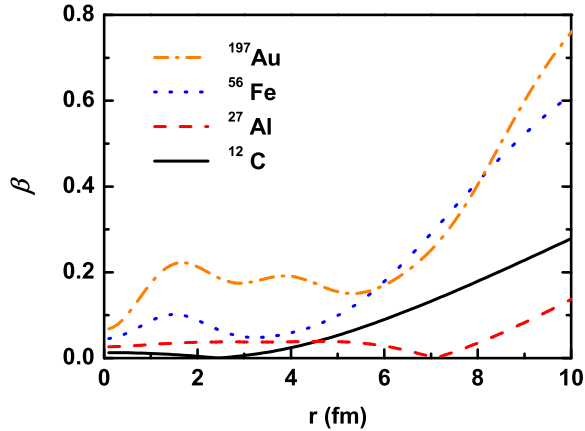}
\end{center}
\caption{Asymmetry distribution for the four nuclei $^{12}$C (a), $^{27}$Al (b), $^{56}$Fe (c) and
$^{197}$Au (d) calculated by the SHF method with the LNS1 parameter set.} \label{fig2}
\end{figure}

To perform the LDA, in Fig.\ref{fig1} we display the neutron,
proton and total density distributions of the four nuclei
$^{12}$C (a), $^{27}$Al (b), $^{56}$Fe (c) and $^{197}$Au (d)
which are calculated by the SHF method.
We adopt the LNS1 parameter set which is obtained by fitting the
properties of asymmetric nuclear matter predicted by the BHF
approach \cite{Gambacurta:2011}. This parameter set gives a satisfying description of properties of finite nuclei, such as binding energy and charge radius. One can find from Fig.\ref{fig1}
that the total densities inside nuclei are not constant along the
radii. For the heaviest nucleus $^{197}$Au, the total density is
found to be nearly constant in the central region and is close to the
saturation density of nuclear matter. Besides, one has to know the
asymmetry distributions inside nuclei for the application of the
LDA. In Fig.\ref{fig2} we display the asymmetry distributions of the
four nuclei $^{12}$C, $^{27}$Al, $^{56}$Fe and $^{197}$Au. The
local asymmetry $\beta(r)$ is given by
$[\rho_n(r)-\rho_p(r)]/[\rho_n(r)+\rho_p(r)]$, where $\rho_n(r)$
and $\rho_p(r)$ represent the neutron and proton densities at
radius $r$, respectively. It is shown in Fig.\ref{fig2} that the
asymmetries inside nuclei are not constant but have fluctuations
along the radii. When the radii are not too large, i.e., the
corresponding densities are not too small, the asymmetries can
be roughly estimated by $(N-Z)/(N+Z)$, where $N$ and $Z$ are the
neutron and proton numbers, respectively. However, the asymmetries
of $^{56}$Fe and $^{197}$Au increase distinctly as the radii
become very large, which is in accordance with the phenomenon of
neutron skin or halo structure in nuclear physics.

One should notice that the LDA will give rise to uncertainties of calculations.
In the recent BHF studies for finite nuclei, Bethe-Goldstone equation is solved directly for finite nuclei and several interesting new results were reported continuously\cite{Liang:2016,Shen:2016,Shen:2017,Shen:2018}. For example, in Ref.\cite{Liang:2016}, it has been shown that the different LDAs generate substantially different results by use of the same Brueckner theory with the same interaction adopted. Therefore, the investigation of the uncertainties caused by the LDA should be concerned in the future research.

\begin{figure}[tbh]
\begin{center}
\includegraphics[width=0.6\textwidth]{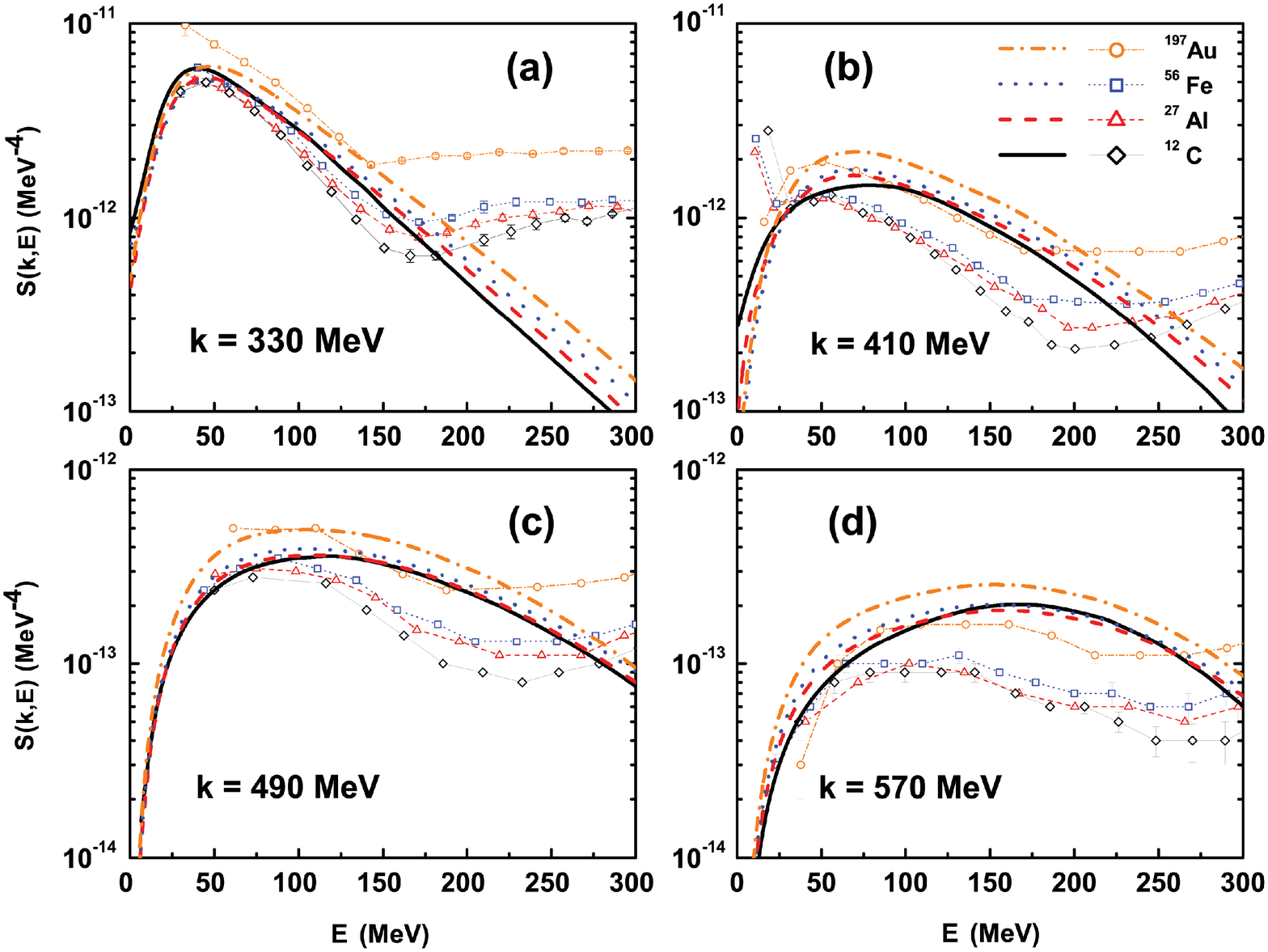}
\end{center}
\caption{By various lines we give spectral functions divided by proton numbers from our calculations for the four nuclei $^{12}$C, $^{27}$Al, $^{56}$Fe and
$^{197}$Au at four different momenta $330$ MeV (a), $410$ MeV (b), $490$ MeV (c) and $570$ MeV (d). Open-symbol dots correspond to the experimental results taken from Ref.\cite{Rohe:2003}.} \label{fig3}
\end{figure}

From Fig.\ref{fig1} and Fig.\ref{fig2} one can know the density and asymmetry at
radius $r$ and consequently obtain the proton spectral function based on Eq.(\ref{eq:specnuclei}) for every nucleus
discussed above. In Fig.\ref{fig3}, we display the spectral functions divided by
the proton numbers for the four nuclei $^{12}$C, $^{27}$Al, $^{56}$Fe and
$^{197}$Au at four different momenta $330$ MeV (a), $410$ MeV (b), $490$ MeV (c) and $570$ MeV (d). Besides, we also show the experimental results via (e, e'p)
at JLab\cite{Rohe:2003} for comparison. One may notice that at low energies the present calculation can well reproduce
the experimental result at small momenta, while the discrepancy between them becomes apparent
with increasing the momentum since the effects of final state interaction (FSI) become significant at high momenta\cite{Frankfurt:2008,benmokhtar:2005,Barbieri:2005}. To include the  FSI in future calculations is potentially able to eliminate the discrepancy between the theoretical and experimental results at high momentum. It is also shown in Fig.\ref{fig3} that the spectral functions from the present calculation decrease with missing energy while the ones from the experiment behave differently.  This can be explained by the existence of the $\Delta$-resonance in the experiment which is not considered in our calculation. The $\Delta$-resonance appears at high missing energy and consequently enhances the spectral function, for example,
sizable contributions from the excitation of the $\Delta$-resonance enhance the spectral function
of $^{12}$C at missing energies above $150$ MeV. The improvement of the calculated spectral functions at high missing energies is expected by inclusion of the $\Delta$-resonance in future theoretical calculations.

In Ref.~\cite{Bozek:2004}, the CD-Bonn interaction was used in the calculation and it is stated that to adopt a nuclear potential with a hard core could bring the calculation closer to the experimental results. In our calculation, the AV18 interaction is adopted, which introduces stronger short-range correlations than the CD-Bonn interaction. Besides, a microscopic TBF is included in our calculation, which also induces strong short-range correlations. Consequently, we find that the spectral functions given by the AV18 interaction are larger than those given by the CD-Bonn interaction and our calculation gives a better description of the experimental results at low momentum. However, the present calculation still cannot give a quantitative
description of the experimental results. To quantitatively reproduce the experimental results is still challenging for nuclear theory.

\begin{figure}[tbh]
\begin{center}
\includegraphics[width=0.5\textwidth]{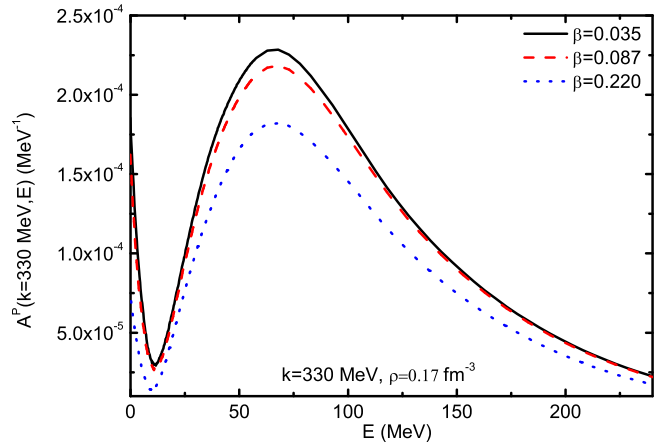}
\end{center}
\caption{Proton spectral function as a function of missing energy
with momentum $k=330$ MeV for three different asymmetries
($\beta=0.035, 0.087$ and $0.22$) at the saturation density in
nuclear matter..} \label{fig4}
\end{figure}
By comparing the four panels (a), (b), (c) and (d) in Fig.\ref{fig3},
one can notice that the spectral function decreases with
momentum for the same nucleus, which means that the possibility to
find a nucleon inside a nucleus decreases by increasing the
momentum. At a given momentum, one may notice from
both the experimental results and our present calculations that the off-shell
spectral function per proton increases when going from $^{12}$C to
$^{197}$Au which is called the target dependence of the scaled proton spectral functions in finite nuclei. To understand this phenomenon, we start the discussion from the proton
spectral function in nuclear matter.

Firstly, we show the proton
spectral function in asymmetric nuclear matter as a function of
the missing energy for three different asymmetries($\beta=0.035,
0.087$ and $0.22$) at the saturation density in Fig.\ref{fig4}.
One can see from Fig.\ref{fig4} that the proton spectral function
decreases with asymmetry of nuclear matter, which means that the
probability to find a proton inside a nucleus decreases with
asymmetry. This result is in accordance with our previous
calculation~\cite{Yin:2013} and can be explained by the tensor
component of nucleon-nucleon
interaction~\cite{Yin:2017,rios:2009}. However, one should notice
that this asymmetry dependence is very weak. From Fig.\ref{fig2}
we can find that the asymmetries of the central regions for these
4 nuclei are less than $0.22$. Thus one can conclude that the
asymmetry distributions inside finite nuclei has negligible
effects on the target dependence of the scaled proton spectral
functions of nuclei.

\begin{figure}[tbh]
\begin{center}
\includegraphics[width=0.5\textwidth]{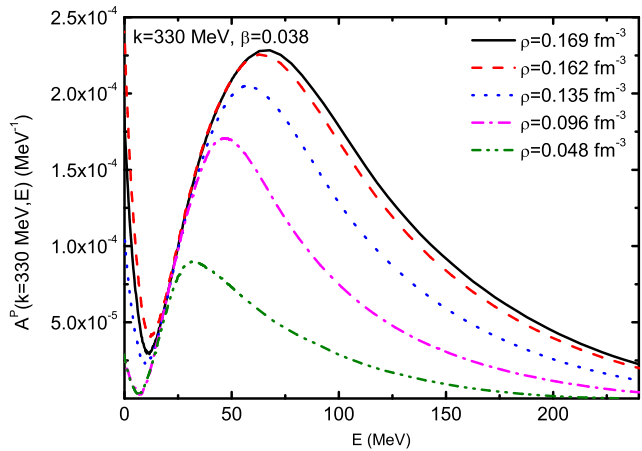}
\end{center}
\caption{Proton spectral function as a function of missing energy
with momentum $k=330$ MeV for five different densities ($\rho=0.169,
0.162, 0.135, 0.096$ and $0.048$ fm$^{-3}$) in nuclear matter with
asymmetry $\beta=0.038$ .} \label{fig5}
\end{figure}
Secondly, in Fig.\ref{fig5} we display the proton spectral function versus the missing energy at five different densities ($\rho=0.169, 0.162, 0.135,
0.096$ and $0.048$ fm$^{-3}$) in asymmetric nuclear matter at a
fixed asymmetry of $\beta=0.038$. The proton
spectral function increases with density of nuclear matter since
higher density induces stronger short-range correlations. One can expect from Eq.(\ref{eq:specnuclei}) that the nuclear spectral function increases with the range of the high density region in a nucleus (one should note that what are plotted in Fig.\ref{fig5} is for the total density and the proton density dependence is expected to have the similar behavior).  Furthermore, according to Eq.(\ref{eq:specnuclei}), the
scaled proton spectral functions of finite nuclei can be written as,
\begin{equation}\label{eq:SSZ}
    \frac{S^p(k,E)}{Z}\propto \frac{\int A^p_h[\rho(r),\beta(r);k,E]d^3\vec r}{\int \rho_{p}(r)d^3\vec
    r}.
\end{equation}
Based on the above discussion, we can find that for a given nucleus, the right hand of Eq.(\ref{eq:SSZ}) is mainly contributed by the proportion of the high density region of the proton. It implies a larger proportion of the high density region of the proton will generate a higher value of the scaled proton spectral function.

\begin{table}[ht]
\caption{Proportions of the high density region of the proton, i.e. $\chi$, and scaled proportions of the high density region of the proton, i.e. $\chi_C$, for the four different nuclei $^{12}$C, $^{27}$Al, $^{56}$Fe and $^{197}$Au. }\label{tab1}
\begin{ruledtabular}
\begin{tabular}{ccc}
 Nuclei  & $\chi$  & $\chi_C$  \\
\hline\\
$^{12}$C & 0.45 & 1.00  \\
$^{27}$Al & 0.63 & 1.39  \\
$^{56}$Fe & 0.68 & 1.49  \\
$^{197}$Au & 0.76 & 1.67  \\
\end{tabular}
\end{ruledtabular}
\end{table}
To  be more explicit, we define a quantity $\chi=\frac{\int_{0}^{r_{h}}
\rho_{p}(r)d^3\vec r}{\int \rho_{p}(r)d^3\vec r}$, where $r_{h}$
represents the range of the high density region of the proton, to signify the proportion of the high density region of the proton. In practice, $r_{h}$ is defined according to the half value of the maximum proton density, i.e., $\rho_{p}(r_{h})=\frac{\rho_{p}^{max}}{2}$. In Tab.\ref{tab1}, in the second column we display $\chi$ for the four different targets considered in the present paper, i.e. $^{12}$C, $^{27}$Al, $^{56}$Fe and $^{197}$Au. In the third column of the Tab.\ref{tab1}, we show the scaled proportions of the high density region of the proton $\chi_C$, which is defined by $\chi_C=\frac{\chi_i}{\chi_{^{12}C}}$ where $i$ runs over $^{12}$C, $^{27}$Al, $^{56}$Fe and $^{197}$Au. From Tab.\ref{tab1} we can see that the proportion of the high density region of the proton increases significantly going from $^{12}$C to $^{197}$Au, which is in accordance with the target-dependent behavior of the scaled proton spectral function in Fig.\ref{fig3}. The role of $\chi$ in the target dependence of the scaled proton spectral function is more pronounced than the one of $\frac{N+Z}{Z}$, which was stressed in Ref.~\cite{Bozek:2004}. We conclude that the proportion of the high density region of the proton plays a crucial role in the target-dependent behavior of the scaled proton spectral function.

\section{Summary}

In summary, we have calculated the proton spectral functions for
finite nuclei by the local density approximation where the nuclear structure is calculated from
the Skyrme-Hartree-Fock method and the nuclear matter calculation is performed
within the framework of the extended Brueckner-Hartree-Fock approach by adopting the
AV18 two-body interaction supplemented with a microscopic three-body force.
Our calculation is in good accordance with the experimental
results at small momenta. Besides, the absence of the
$\Delta$-resonance in our calculation leads to a discrepancy with
the experimental results at high missing energies. From our calculation,
we also find that the scaled proton spectral function increases
when going from $^{12}$C to $^{197}$Au, which is consistent with
the experimental results. In asymmetric nuclear matter, the effects of short-range correlations increase with density and hence the spectral function becomes larger at higher densities. By a further investigation, we find that the proportion of the high density region of proton plays a significant role in the target dependence of the scaled proton spectral function. Inclusion of the
$\Delta$-resonance in nucleon-nucleon interaction as well as the final state interaction are
expected to improve our results in future
calculations~\cite{Piarulli}. In addition, the effects of the local density approximation on spectral functions of finite nuclei should also be investigated in the following research.
% is shown to result in
%occupation of its hole states while
%depletion below its Fermi
%momentum becomes smaller, while the proton depletion below its Fermi momentum
% becomes larger,

%The density dependence and isospin-asymmetry dependence of the nucleon momentum
%distributions have been predicted. The density dependence, as a function of the ratio $k_F/k$,
%turns out to be rather weak , in agreement with the results given in Ref.[6,7].
%The TBF effect turns out to be negligibly small
%  at relatively low densities around and below the saturation density. Whereas, at high densities, the TBF
%  affects the neutron and proton momentum distributions sizably. It may lead to

%  We have also calculated the
%   quasi-particle strength at the Fermi surface. The obtained result of    is close
%   to the value 0.63 obtained in NIKHEF experiment for P-shell orbitals [16] and in
%   good agreement with experimental data [4,8].

\section*{Acknowledgments}

{The authors appreciate valuable suggestions from Dr. Daniela Kiselev (-Rohe). Peng Yin appreciates the kind hospitality of Professor James P. Vary during his visit to Iowa State University. The work is supported by the National Natural Science Foundation of China (11435014, 11175219, 11373038, 11603046, 11705240, 11505241), the 973 Program of China (Grant No. 2013CB834405), the Strategic Priority Research Program ¡°The Emergence of Cosmological Structures¡± of the Chinese Academy of Sciences, Grant No. XDB09000000, National Key R\&D Program of China No. 2017YFA0402600, Guizhou Provincial Key Laboratory of Radio Astronomy and Data Processing, Guizhou Normal University, Guiyang 550001, China
 and the Knowledge Innovation Project (KJCX2-EW-N01) of the Chinese Academy of Sciences. This work is also partially supported by the CUSTIPEN (China-U.S. Theory Institute for Physics with Exotic Nuclei) funded by the U.S. Department of Energy, Office of Science under Grant No. DE-SC0009971.}

\end{document}